\documentclass[twocolumn,superscriptaddress,aps,showpacs,amsmath,amstex,amssymb,citeautoscript,floatfix,prl,preprintnumbers]{revtex4-2}
\usepackage{graphicx}
\usepackage{latexsym}
\usepackage{amssymb}
\usepackage{amsmath}
\usepackage{amsfonts}
\usepackage{upgreek}
\usepackage{subfigure}
\usepackage{bm}
\usepackage{nicefrac}
\usepackage{bbold}
\usepackage{aliascnt}
\usepackage{amsthm}
\usepackage{hyperref}
\usepackage{cleveref}
\usepackage{verbatim}
\usepackage{multirow}
\usepackage{color}
\usepackage{comment}
\usepackage{xcolor}
\usepackage{soul}
\usepackage{tikz}

\usepackage{braket}

\newcommand{\bbrakket}[1]{\langle\!\langle #1\rangle\!\rangle}
\newcommand{\eqs}[1]{\begin{equation}\begin{split}#1\end{split}\end{equation}}

\theoremstyle{plain}

\crefname{theorem}{theorem}{theorems}
\Crefname{theorem}{Theorem}{Theorems}

\newaliascnt{lemma}{theorem}

\aliascntresetthe{lemma}
\crefname{lemma}{lemma}{lemmas}
\Crefname{lemma}{Lemma}{Lemmas}

\newaliascnt{proposition}{theorem}

\aliascntresetthe{proposition}
\crefname{proposition}{proposition}{propositions}
\Crefname{proposition}{Proposition}{Propositions}

\theoremstyle{definition}
\newaliascnt{definition}{theorem}

\aliascntresetthe{definition}
\crefname{definition}{definition}{definitions}
\Crefname{definition}{Definition}{Definitions}

\newaliascnt{assumption}{theorem}

\aliascntresetthe{assumption}
\crefname{assumption}{assumption}{assumptions}
\Crefname{assumption}{Assumption}{Assumptions}

\newaliascnt{goal}{theorem}

\aliascntresetthe{goal}
\crefname{goal}{goal}{goals}
\Crefname{goal}{Goal}{Goals}

\newaliascnt{example}{theorem}

\aliascntresetthe{example}
\crefname{example}{example}{examples}
\Crefname{example}{Example}{Examples}

\newcommand{\Harvard}{Department of Physics, Harvard University, Cambridge, MA 02138, USA}
\newcommand{\ITAMP}{ITAMP, Center for Astrophysics $\vert$ Harvard \& Smithsonian, Cambridge, MA 02138, USA}

\begin{document}
\title{
Model-agnostic cooling algorithms for strongly interacting fermions
}

\author{Henning Schl\"omer}
\altaffiliation{Corresp. author: \href{mailto:henning.schloemer@cfa.harvard.edu}{henning.schloemer@cfa.harvard.edu}}
\affiliation{\ITAMP}
\affiliation{\Harvard}

\author{Liyuan Chen}
\affiliation{\Harvard}
\author{Susanne F. Yelin}
\affiliation{\Harvard}
\author{Hong-Ye Hu}
\altaffiliation{Corresp. author: \href{mailto:hongyehu.physics@gmail.com}{hongyehu.physics@gmail.com}}
\affiliation{\Harvard}

\begin{abstract}
Strongly interacting fermions underpin some of the most challenging problems in condensed matter physics, such as high-temperature superconductivity. The low-energy states of these systems encode their essential microscopic properties, yet remain largely inaccessible to classical methods. Quantum simulation offers a promising path forward, and among state-preparation strategies, engineered dissipation has emerged as a particularly compelling approach. Existing cooling protocols, however, typically rely on knowledge of the quasiparticle spectrum or mappings to free-fermion limits. In this letter, we introduce a randomized, symmetry-preserving cooling algorithm that requires no spectral information, using only local coupling operators to ancilla degrees of freedom with randomly sampled energy splittings to drive generic fermionic systems toward their low-energy manifold. We benchmark the protocol on canonical correlated fermionic models relevant to high-temperature superconductors, spanning metallic, density-wave, paired, superconducting, and phase-separated phases. Across all models, we observe universal cooling behavior: monotonic energy relaxation, concentration of spectral weight at low energies, and stabilization of correlated ground-state order. Our results establish randomized dissipative cooling as a general strategy for preparing strongly correlated fermionic states on programmable quantum devices.
\end{abstract}

\maketitle

\textit{Introduction.---}Understanding strongly interacting fermionic systems remains one of the central challenges in modern physics. From high temperature superconductors to correlated molecular systems, the interplay of charge, spin, and orbital degrees of freedom gives rise to rich phase diagrams featuring magnetism, charge order, and unconventional superconductivity~\cite{Altman2021, Preskill2018}. However, reaching low-energy states of these systems is difficult: classical methods face fundamental obstacles such as the fermion sign problem in quantum Monte Carlo and rapid entanglement growth, which limit tensor network approaches beyond one dimension~\cite{SchollwoeckDMRG, SchollwoeckDMRG2, WhiteDMRG, Wecker2015_2}.

Quantum simulation offers a promising route to overcome these limitations, and significant progress has been made on both analog and digital platforms. Unitary state-preparation strategies, including adiabatic~\cite{Born1928} and counter-diabatic~\cite{Dries2017,Claeys2019,Odelin2019,Gjonbalaj2025,Rudi2025} protocols, steering methods~\cite{Roy2020,Herasymenko2023,Volya2024}, and variational~\cite{Wecker2015,Wecker2015_2, Cerezo2021} or learning-based~\cite{Yao2021,Schloemer2025ML} schemes, have been explored extensively. In parallel, engineered dissipation has emerged as a compelling alternative, in which energy and entropy are removed by coupling the system to auxiliary degrees of freedom, so that the target state emerges as the steady state of open-system dynamics~\cite{Kraus2008, Diel2008, Verstraete2009, Tindall2019, Neri2025, Lin2025, Westhoff2025}. Utilizing such open quantum dynamics potentially can overcome certain challenges of unitary state preparations~\cite{PRXQuantum.6.010329}. From a related perspective, algorithmic cooling protocols based on repeated system-ancilla interactions extract entropy to the environment cycle-by-cycle~\cite{Polla2021, Mi2024,Molpeceres2025,Langbehn2025, Marti2025}. If the target models admit a well-defined quasiparticle description, kinetic theories can analytically characterize cooling efficiency~\cite{Lloyd2025}, and conditions for rapid convergence have been established in quasi-free settings~\cite{Zhan2025}.

\begin{figure}[t!]
\includegraphics[width=\columnwidth]{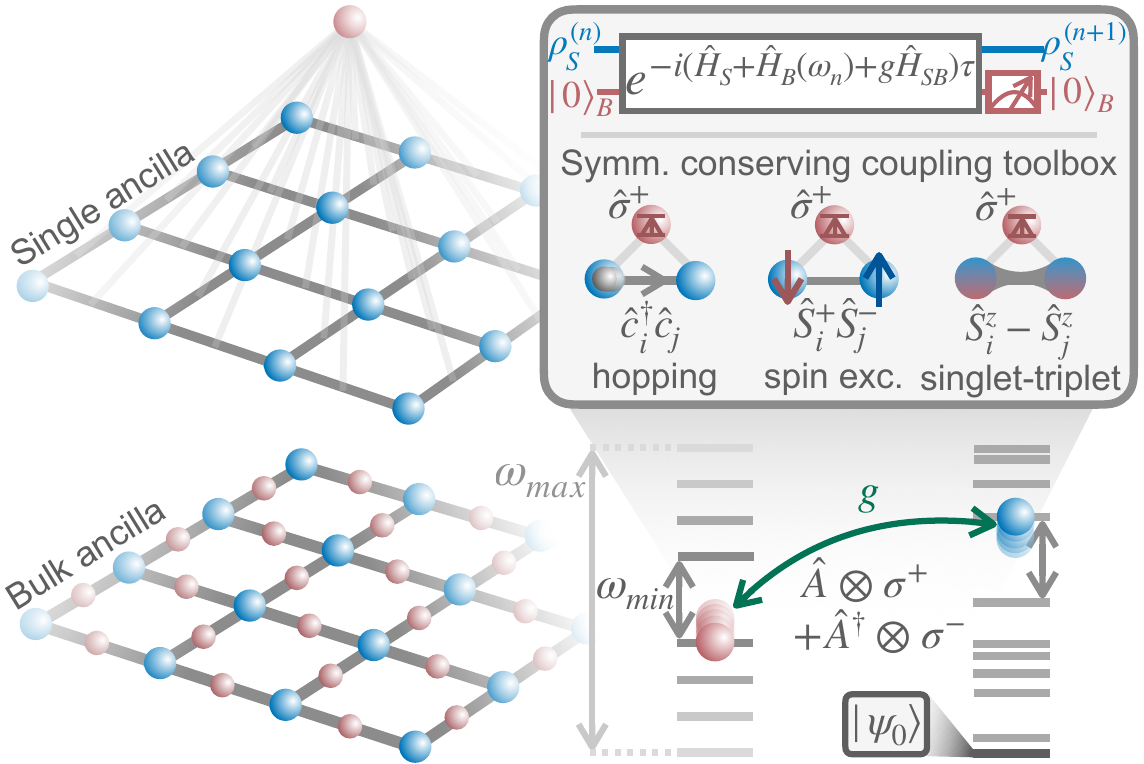}
\caption{\textbf{Cooling algorithm for strongly interacting fermions.} Schematic illustration of the algorithm using either a single ancilla or a bulk ancilla architecture to mediate energy-selective interactions with the system. A coupling toolbox enables cooling within a chosen symmetry sector. Energy-selective interactions between system and ancillas with randomly chosen energy splittings transfer population from higher to lower many-body energy levels.}
\label{fig:fig0}
\end{figure}

On the experimental side, analog quantum gas microscopes have achieved impressive control over Fermi-Hubbard-type models through carefully engineered adiabatic ramps~\cite{Bloch2008, Esslinger2010, Bernien2017, Bohrdt2020, Xu2025, dqyf-kl8x, PRXQuantum.5.040341}, while digital approaches have so far been largely restricted to shallow variational circuits~\cite{Cade2020, Stanisic2022} or pseudo-adiabatic preparations~\cite{Dreyer_1, Dreyer_2}. Despite this progress, systematically reaching the low-energy states of strongly correlated fermionic models with competing orders remains an open challenge~\cite{Wecker2015_2}.

For weakly interacting fermions, cooling protocols based on system-ancilla couplings have been developed using randomized interactions~\cite{Langbehn2025} or explicit coupling to highly non-local eigenstate operators derived from the free-fermion limit~\cite{Marti2025}. However, it is unclear whether such approaches remain effective in strongly interacting regimes, such as doped Mott insulators, where the low-energy physics is no longer captured by quasi-free fermions.

In this work, we address this question by developing randomized, symmetry-preserving cooling protocols for strongly interacting fermionic systems using only \textit{local} coupling operators. Across a wide range of correlated phases in representative lattice models, we demonstrate that randomized cooling provides a universal and robust strategy for preparing low-energy states of interacting fermionic Hamiltonians.

\textit{Symmetry-preserving cooling algorithm.---}We consider interacting fermionic lattice systems described by Hamiltonians $\hat H_S$ that conserve global $U(1)$ charges, such as total particle number or total spin. In a broad range of applications, one is interested in preparing states within a fixed symmetry sector. This is also standard in state-of-the-art numerical approaches to ground-state problems, where calculations are typically restricted to fixed symmetry sectors, for instance in Hubbard-type models~\cite{SchollwoeckDMRG2}.

The physical fermionic degrees of freedom reside on the lattice sites shown in blue in Fig.~\ref{fig:fig0}. We introduce $N_{\text{anc}}$ ancillary qubits (red in Fig.~\ref{fig:fig0}) that serve as auxiliary degrees of freedom mediating the cooling dynamics. In each cooling cycle, the ancillas are initialized in their ground state, $\hat\rho_{\text{anc}}^{(0)}=\ket{0}\!\bra{0}^{\otimes N_{\text{anc}}}$. System and ancillas are then evolved jointly for a time $\tau$ under the unitary $\hat U(\{\omega\}) = e^{-i\hat H(\{\omega\})\tau}$ generated by
\begin{equation}
\hat H(\{\omega\})
= \hat H_S
+ \sum_{\ell=1}^{N_{\text{anc}}} \hat H_{\text{anc}}^{(\ell)}(\omega_j)
+ g \sum_{j=\ell}^{N_{\text{anc}}} \left( \hat A_\ell \otimes \hat\sigma_\ell^+ + \text{h.c.} \right),
\label{eq:te}
\end{equation}
where $\hat H_{\text{anc}}^{(\ell)}(\omega_\ell) = - \nicefrac{\omega_\ell}{2}\,\hat\sigma_{\ell}^z$ sets the level splitting of the ancilla $\ell$, and $g$ controls the system-ancilla coupling. After the joint evolution, the ancilla degrees of freedom are traced out and reset, yielding a completely positive trace-preserving map from cooling step $n\rightarrow n+1$, 
\begin{equation}
\hat\rho^{(n+1)}
=
\mathrm{Tr}_{\text{anc}}
\!\left[
\hat U(\{\omega\})
\left(
\hat\rho^{(n)} \otimes \hat\rho_{\text{anc}}^{(0)}
\right)
\hat U^\dagger(\{\omega\})
\right].
\label{eq:channel}
\end{equation}
The operators $\hat A_\ell$ act solely on the system and determine which transitions the cooling cycle can address. Near solvable limits, such as weakly interacting regimes, a natural choice is to take $\hat A_\ell$ to be a quasiparticle annihilation operator~\cite{Lloyd2025}. In strongly interacting systems, however, no well-defined quasiparticle description exists, and the design of $\hat A_\ell$ becomes a central question. In this work, we show that effective cooling can be achieved by selecting $\hat A_\ell$ from a generic toolbox of \textit{local} operators. Remarkably, the precise form of $\hat A_\ell$ is not critical: we conjecture that whenever the system Hamiltonian satisfies the eigenstate thermalization hypothesis (ETH), the matrix elements $\langle k|\hat A|k'\rangle
$ between many-body eigenstates $\ket{k}$ and $\ket{k'}$ of $\hat{H}_S$ provide sufficient spectral connectivity to ensure energy relaxation within the rotating wave approximation.

Concretely, we select $\hat A_\ell$ from a minimal symmetry-preserving coupling toolbox: directed hopping $\hat A^c_{ij}=\hat c_i^\dagger \hat c_j$, spin-exchange $\hat A^{\mathrm{exc}}_{ij}=\hat S_i^+ \hat S_j^-$, and singlet-triplet switching $\hat A^{st}_{ij}=\hat S_i^z-\hat S_j^z$, each defined on bonds $\ell = \braket{ij}$ of the underlying lattice (additional operators are discussed in the End Matter).
Within this framework, various system-ancilla architectures are possible. We consider two: (i) a single-ancilla protocol, in which an ancilla qubit is coupled to all bonds via $\sum_{ij}\hat A_{ij}$, and (ii) a bulk architecture, in which an individual ancilla is coupled locally to each bond $\langle ij\rangle$ (see Fig.~\ref{fig:fig0}). The single-ancilla setup is more convenient for numerical simulation, while the bulk architecture is better suited to near-term hardware implementations~\cite{majidy2024building}, for instance, neutral atoms in reconfigurable tweezer arrays~\cite{Zhou2025_na, Bluvstein2026} combined with local fermion-to-qubit encoding schemes~\cite{Verstraete_2005, Derby_2021, Jafarizadeh_2025_PRR, Evered2025}.

For weak coupling 
$g$ and sufficiently long interaction times, transitions between many-body eigenstates $\ket{k}$ and $\ket{k'}$ of $\hat{H}_S$ occur at rates proportional to $g^2 \, \big| \langle k | \hat{A} | k' \rangle \big|^2$, subject to an approximate resonance condition $E_k - E_{k'} \simeq \omega$. In this limit, the rotating wave approximation enforces energy conservation of each transition. Crucially, because the ancillas are reset to their ground state after every cycle, only processes that transfer energy from the system to the ancilla can accumulate, while the reverse heating processes are suppressed~\cite{Mi2024,Molpeceres2025,Langbehn2025, Lloyd2025} (see End Matter). The net effect is a biased random walk in energy space that preferentially drives population toward lower-energy states.
\begin{figure}[t!]
\includegraphics[width=\columnwidth]{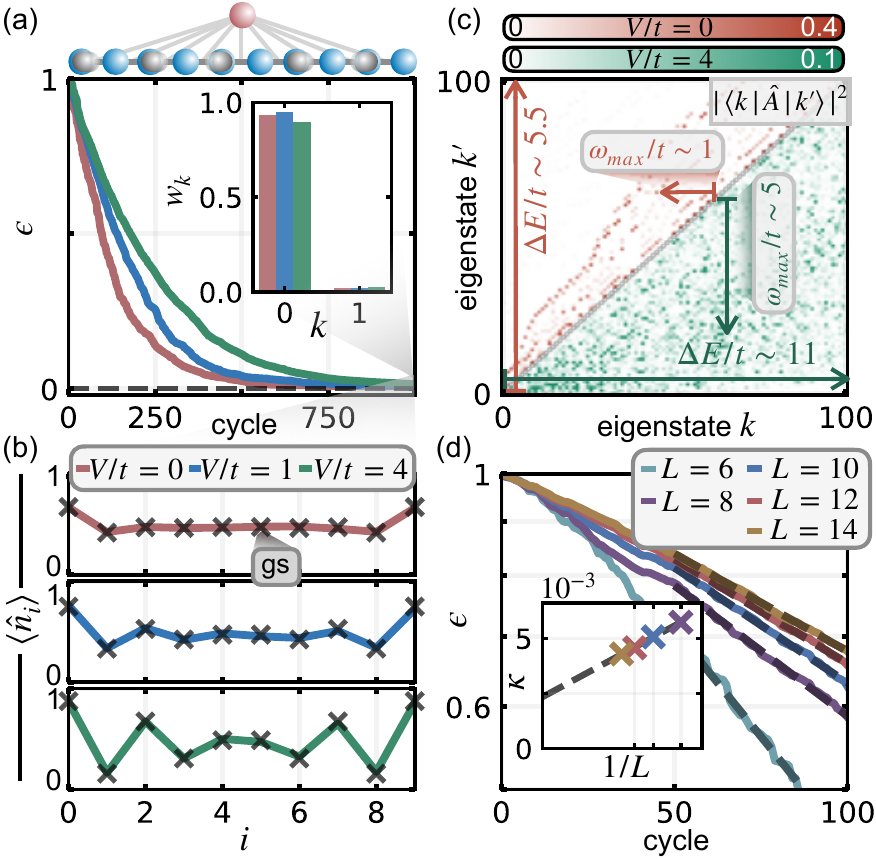}
\caption{\textbf{Cooling dynamics in the spinless $\mathbf{t}$-$\mathbf{t'}$-$\mathbf{V}$ chain.}
(a) Normalized energy density $\varepsilon(n)$, Eq.~\eqref{eq:enden}, vs. cooling cycle $n$ for $t/t'=2$, $g/t=0.025$, $\tau t=40$, $V/t=0,1,4$, $L=10$ sites and $N_p=5$. The initial state has all fermions localized on the left side of the chain. Inset: steady state weights $w_k$ of the lowest eigenstates $k$. (b) Local densities $\langle n_i \rangle$ in the cooled steady state (solid lines) compared with the exact ground state (crosses). (c) Matrix elements $|A_{kk'}|^2$ in the energy eigenbasis for $V/t=0$ (upper part) and $V/t=4$ (lower part), restricted to the lowest 100 eigenstates. Arrows indicate the many-body bandwidth $\Delta E = E_{99}-E_0$ and the cutoff $\omega_{\mathrm{max}}$ corresponding to optimal cooling. (d) Early-time exponential fits $\varepsilon(n) \propto \exp(-\kappa n)$ as a function of system size for $V/t=1$, $\omega_{\text{max}} = 2t$, and other parameters as in (a); fits are shown by dashed lines. The finite-size scaling of $\kappa$ (inset) indicates a finite relaxation rate in the large system limit.}
\label{fig:fig1}
\end{figure}
\begin{figure*}
\includegraphics[width=\textwidth]{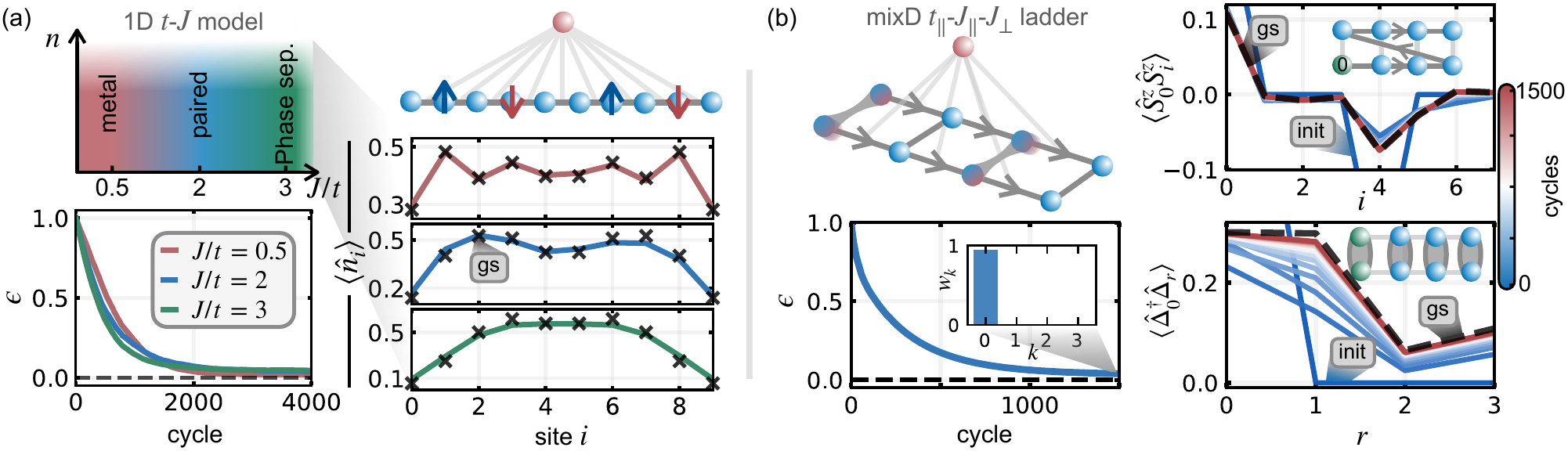}
\caption{\textbf{Cooling doped Mott insulators.}
(a) 1D $t$-$J$ chain across metallic, paired, and phase-separated regimes as $J/t$ is tuned. Charge and spin coupling channels are combined (see main text); $L=10$, $N_p=4$, $g=0.05$, $\tau t=10$, $\omega/t\in[0,3]$, and total spin sector $\hat{S}^z_{\mathrm{tot}}=0$. Shown are normalized energy relaxation for $J/t=0.5,2,3$ and corresponding on-site densities of the steady state (solid lines) and the ground state (black crosses). The initial state is a N\'eel product state with particles sitting at the boundary. 
(b) Mixed-dimensional $t_{\parallel}$-$J_{\parallel}$-$J_{\perp}$ ladder for a $4\times2$ system with $N_p=4$, $J_{\perp}/t_\parallel=2$, $J_{\parallel}/t_\parallel=1$. Cooling is implemented via charge hopping $\hat{A}^c_{\parallel}$ with fixed $\omega/t_\parallel =0.8$, starting in an initial product state of rung-singlets and hole pairs. Energy relaxation, steady-state decomposition, spin-spin correlations $\langle \hat{S}_0^z \hat{S}_i^z \rangle$, and pair-pair correlations $\langle \hat{\Delta}_0^\dagger \hat{\Delta}_r \rangle$ are shown as a function of the protocol cycle; black dashed lines denote ground-state values.}
\label{fig:fig2}
\end{figure*}
To avoid relying on spectral knowledge, the ancilla splittings $\omega_\ell$ are drawn randomly from a tunable interval $\omega_\ell \in [\omega_{\text{min}},\omega_{\text{max}}]$ in each cycle~\cite{Langbehn2025}.
Over many repetitions, this covers a broad range of many-body energy differences, enabling cooling without fine-tuning to specific transitions. 

The full protocol reduces to a simple recipe: (i) fix the interacting fermionic Hamiltonian $\hat H_S$ and the relevant $U(1)$ sector, (ii) choose a coupling operator $\hat A$ from the local toolbox, and (iii) iterate the cooling cycle with randomly sampled ancilla frequencies. Provided the chosen coupling induces sufficient connectivity in the many-body eigenbasis, this procedure generically drives the system toward its low-energy states. In the following, we benchmark this approach on three canonical strongly interacting fermion models, all of which are motivated by the physics of cuprate and nickelate high-$T_c$ superconductors.

\textit{Spinless fermions.---}We begin by considering interacting spinless fermions on a one-dimensional (1D) lattice of length $L$ at fixed particle number $N_p$. We include NN hopping $t$, NNN hopping $t'$, and a NN interaction $V$,
\begin{equation}
\begin{aligned}
\hat{H}_{tt'V} = - t \sum_{\braket{ij}} ( \hat{c}_i^\dagger \hat{c}_{j} + \text{h.c.} ) - t' &\sum_{\bbrakket{ij}} ( \hat{c}_i^\dagger \hat{c}_{j} + \text{h.c.}) \\ 
&+ V \sum_{i} \hat{n}_i \hat{n}_{i+1},
\end{aligned}
\end{equation}
with $\hat{n}_i = \hat{c}_i^\dagger \hat{c}_i$. For $t'=0$ this reduces to the standard $t$-$V$ chain, which has a transition from a metallic Luttinger liquid phase at weak $V/t$ to a charge-density-wave phase at strong repulsion~\cite{Giamarchi2003}. The inclusion of $t'$ breaks simple mappings to local spin systems and enhances genuine fermionic correlations.

We benchmark the cooling protocol using density-matrix calculations in a fixed particle-number sector and monitor the normalized energy density 
\begin{equation}
\varepsilon(n) = \frac{E(n)-E_{\mathrm{gs}}}{E(0)-E_{\mathrm{gs}}},
\label{eq:enden}
\end{equation}
where $E(n)=\mathrm{Tr}[\hat{H}\hat{\rho}^{(n)}]$ is the energy after $n$ cooling cycles. By construction, $\varepsilon=0$ corresponds to the exact ground-state energy. Fig.~\ref{fig:fig1}~(a) shows $\varepsilon(n)$ for interaction strengths $V/t=0,2,4$ for the single-ancilla protocol ($N_{\text{anc}} = 1$) at system size $L=10$, $N_p=5$, and a hopping coupler $\sum_{\braket{ij}}\hat{A}^c_{ij}$. For all interactions the energy decreases monotonically over the course of the cooling cycles, indicating systematic removal of high-energy weight and reaching high ground-state fidelities after around 1000 cycles (inset). Local density profiles $\langle \hat{n}_i\rangle$ in the final steady state are compared to exact ground-state densities in Fig.~\ref{fig:fig1}~(b). In all cases the cooled state reproduces the ground-state density pattern with high accuracy, showing that the algorithm captures the qualitative change in order across the interaction-driven transition. 

Although we here choose a specific ordering of site pairs in the hopping coupler, we have verified that the performance remains essentially unchanged when using a coupler defined along a randomly chosen path (see End Matter). This indicates that the precise microscopic realization of the coupling operator does not significantly affect the performance, and that the protocol yields generic cooling behavior.

To understand the mechanism at the spectral level, Fig.~\ref{fig:fig1}~(c) shows the matrix elements $|A_{kk'}|^2 = |\langle k|\hat{A}|k'\rangle|^2$ of the first 100 eigenstates of $\hat{H}_{tt'V}$ for $V/t=0$ and $V/t=4$. For the non-interacting case, most of the weight is concentrated along narrow bands corresponding to low-energy particle-hole excitations. In this regime, bath frequencies of order $\omega \in [0,t]$ are most effective, as indicated by the arrows.
In contrast, at $V/t=4$ the matrix elements of $\hat{A}$ are distributed more broadly across the eigenbasis, however with smaller overall amplitude $|A_{kk'}|^2$. In this case, most effective cooling can be achieved by choosing larger bath frequencies $\omega \in [0,5t]$.

Fig.~\ref{fig:fig1}~(d) analyzes the scalability of the protocol with increasing system size. We extract the early-time relaxation rate by fitting the normalized energy to an exponential form, $\epsilon(n) \propto \exp[-\kappa n]$.
At half-filling with fixed $t/t'=2$, $V/t=1$ and $\omega \in [0,2t]$, a scaling analysis suggests that $\kappa$ approaches a finite value as $L$ increases. This indicates that the rapid early-time energy relaxation remains effective in large systems. We emphasize, however, that this does not imply a system-size-independent thermalization or convergence time: the approach to the steady state may still slow down with increasing system size, for instance due to a closing Lindbladian gap.

\textit{Doped Mott insulators.---}We now turn to strongly correlated fermionic systems where charge and spin degrees of freedom are intrinsically intertwined. In particular, we consider state preparation of the 1D $t$-$J$ Hamiltonian
\begin{equation}
\hat{H}_{tJ} =
- t \sum_{\langle ij \rangle,\sigma}
\left( \hat{\tilde{c}}_{i\sigma}^\dagger \hat{\tilde{c}}_{j\sigma} + \text{h.c.} \right)
+ J \sum_{\langle ij \rangle}
\left( \hat{\mathbf{S}}_i \cdot \hat{\mathbf{S}}_j
- \frac{1}{4} \hat{n}_i \hat{n}_j \right),
\end{equation}
where $\hat{\tilde{c}}_{i\sigma}$ are fermionic operators acting in the projected Hilbert space without double occupancy, $\hat{\mathbf{S}}_i$ denotes the local spin operator, and $\hat{n}_i$ the density operator. The ratio $J/t$ controls the competition between kinetic delocalization and magnetic exchange. Upon tuning $J/t$, the ground state exhibits qualitatively distinct regimes: a metallic phase at small $J/t$, a paired regime with dominant superconducting correlations at intermediate $J/t$, and phase separation at large $J/t$~\cite{Ogata1991, Moreno2011}, see Fig.~\ref{fig:fig2}~(a). 

To allow relaxation in both charge and spin sectors, we couple NN and NNN hopping and spin-exchange channels, 
$\hat{A} = \alpha \hat{A}^c + \beta \hat{A}^{\mathrm{exc}}$~\footnote{
Here $\hat{A}^c = \sum_{\langle ij\rangle, \sigma} \hat c_{i,\sigma}^\dagger \hat c_{j,\sigma} 
+ \sum_{\langle\!\langle ij\rangle\!\rangle, \sigma} \hat c_{i,\sigma}^\dagger \hat c_{j,\sigma}$ 
denotes directed hopping including both nearest-neighbor $\langle ij\rangle$ 
and next-nearest-neighbor $\langle\!\langle ij\rangle\!\rangle$ bonds on the 1D lattice. 
Similarly, the spin-exchange operator is 
$\hat{A}^{\mathrm{exc}} = 
\sum_{\langle ij\rangle} \hat S_i^+ \hat S_j^- 
+ \sum_{\langle\!\langle ij\rangle\!\rangle} \hat S_i^+ \hat S_j^-$.}. This way, cooling dynamics can connect states that differ in charge configuration as well as in spin alignment; by doing a scan over $\alpha$ and $\beta$, we find best cooling results for $\alpha =0.4$ and $\beta = 0.6$ (see End Matter). During cooling, the normalized energy $\varepsilon$ decreases in a nearly identical fashion for $J/t=0.5,2,3$, shown in Fig.~\ref{fig:fig2}~(a).

The on-site densities $\langle \hat{n}_i \rangle$ provide a direct physical signature of successful cooling. In the metallic regime, the final state exhibits $N_p$ maxima distributed along the chain. In the paired regime, the density profile reorganizes into $N_p/2$ peaks, consistent with the formation of bound pairs. In the phase-separated regime, particles cluster into a single domain, leading to a broad density peak in the center of the open chain~\cite{Moreno2011}. In all three cases, the cooled steady state reproduces the qualitative ground-state density structure, indicating that the cooling mechanism does not rely on a particular phase structure or gap property, but instead exploits generic spectral connectivity in the many-body eigenbasis.

We next consider a mixed-dimensional ladder described by
\begin{equation}
\begin{aligned}
H_{t_{\parallel} J_{\parallel} J_{\perp}} =
- t_{\parallel} &\sum_{\langle ij \rangle_{\parallel},\sigma}
\left( \hat{\tilde{c}}_{i\sigma}^\dagger \hat{\tilde{c}}_{j\sigma} + \text{h.c.} \right) \\ 
&+ \sum_{\delta = \parallel, \perp} \sum_{\langle ij \rangle_{\delta}} J_{\delta}
\left( \hat{\mathbf{S}}_i \cdot \hat{\mathbf{S}}_j
- \frac{1}{4} \hat{n}_i \hat{n}_j \right),
\end{aligned}
\end{equation}
where $\langle ij \rangle_{\parallel}$ and $\langle ij \rangle_{\perp}$ denote bonds along the legs and rungs of the ladder, respectively, see Fig.~\ref{fig:fig2}~(b). This model provides a minimal framework for the description of bilayer nickelate superconductors under pressure~\cite{Sun2023,Zhang2024,Oh2023,Qu2023,Schloemer2024_LNO}, and has attracted significant interest in the context of quantum simulation of high-temperature superconductivity~\cite{Hirthe2023,Schloemer2024_OL,Dreyer_2}. In the model, rung exchange $J_{\perp}$ favors singlet formation; parallel hopping leads to these pairs becoming coherent, resulting in an $s$-wave superconducting state in the doped system~\cite{Schloemer2024_LNO}.

We start the cooling protocol from a product state composed of alternating singlets and hole-hole pairs, corresponding to the particle density expected in bilayer nickelates~\cite{Sun2023}. Cooling is implemented via directed hopping along the ladder legs, $\hat{A}^c_{\parallel}$, without additional spin channels, and we here fix the ancilla frequency to a value around the hopping scale, $\omega/t_{\parallel} = 0.8$ (see End Matter for a more detailed discussion). As shown in Fig.~\ref{fig:fig2}, the system's state $\hat{\rho}_S$ relaxes to a steady state with nearly unit weight in the target ground state. Since the initial state already contains local singlets, cooling primarily requires lowering their kinetic energy and establishing phase coherence along the ladder, suggesting why a purely charge-based channel suffices to achieve efficient cooling.

The build-up of correlations is illustrated by the evolution of spin-spin correlations $\langle \hat{S}_0^z \hat{S}_i^z \rangle$ and coherent pair-pair correlations $\langle \hat{\Delta}_0^\dagger \hat{\Delta}_r \rangle$, where the rung singlet creation operator is defined as 
$\hat{\Delta}_{r}^\dagger=\nicefrac{1}{\sqrt{2}} (
\hat{\tilde{c}}_{i,\uparrow}^\dagger \hat{\tilde{c}}_{j,\downarrow}^\dagger- \hat{\tilde{c}}_{i,\downarrow}^\dagger \hat{\tilde{c}}_{j,\uparrow}^\dagger)$,
creating a spin singlet across a rung bond $r=\langle i,j\rangle_\perp$. As cooling proceeds, spin correlations relax toward their ground-state profile, while superconducting pair correlations are dynamically built up and converge to the exact values.

\begin{figure}[t!]
\includegraphics[width=\columnwidth]{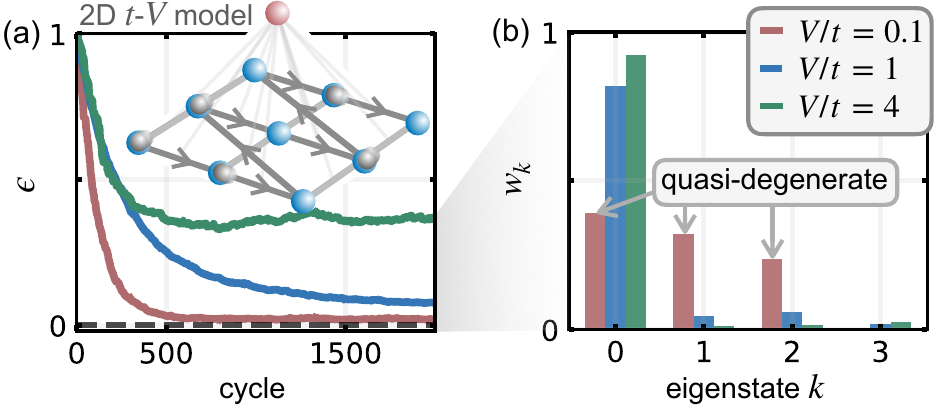}
\caption{\textbf{Spinless fermions in 2D.}
Cooling dynamics (a) and steady state decomposition (b) of interacting spinless fermions on a $3\times 3$ lattice with $N_p = 5$, $g/t=0.05$, $\tau t = 10$, and $\omega_{\text{max}}/t = 2,6,8$ for $V/t = 0.1,1,4$, respectively. The coupling operator is chosen as a directed $Z$-shaped path in the 2D lattice, shown in the inset of panel (a).}
\label{fig:fig3}
\end{figure}

\textit{Towards two dimensions.---}To demonstrate that our protocol extends beyond one dimension, we consider interacting spinless fermions on a $3\times 3$ lattice with $N_p=5$ particles with NN hopping $t$ and interactions $V$. A single ancilla is coupled to the system via a directed hopping operator $\hat{A}^c$ that traverses the lattice along a $Z$-shaped path, see Fig.~\ref{fig:fig3}~(a).

In the weakly interacting regime the three lowest eigenstates are quasi-degenerate, becoming exactly degenerate for $V/t=0$~\footnote{For the non-interacting spinless fermion problem on a \(3\times 3\) square lattice with open boundary conditions, the nearest-neighbor hopping Hamiltonian has single-particle energies
\(\epsilon_{n_x,n_y}=-2t[\cos(\pi n_x/4)+\cos(\pi n_y/4)]\), with \(n_x,n_y=1,2,3\). The corresponding spectrum has degeneracies \(1,2,3,2,1\) at energies \(-2\sqrt{2}t\), \(-\sqrt{2}t\), \(0\), \(+\sqrt{2}t\), and \(+2\sqrt{2}t\), respectively. For \(N=5\) spinless fermions, the lowest three orbitals are fully occupied and the remaining two fermions occupy the threefold-degenerate zero-energy shell. Hence the many-body ground-state degeneracy is \(\binom{3}{2}=3\).}. As seen by the decomposition of the steady state in Fig.~\ref{fig:fig3}~(a), the protocol results in similar weight within this low energy manifold. For stronger interactions the low-energy spectrum separates, and a sizable many-body gap opens above the ground state, leading to high fidelities after approaching a steady state. We note that in this case, $\epsilon$ does not approach zero, which can be attributed to the large many-body gap: even small populations of excited states in the final state lead to significant shifts of the expectation value of the energy.

Beyond the settings discussed above, we assess two experimentally relevant aspects in the End Matter. First, we demonstrate that the cooling dynamics is robust against moderate global dephasing noise. Second, we show that the system-ancilla coupling architecture does not critically affect cooling performance: global and bond-resolved bulk couplings yield comparable results, while boundary-only cooling proves less efficient. Together, these findings suggest realistic pathways toward implementing the proposed cooling algorithm on near-term hardware, which we will explore in detail in an upcoming work.

\textit{Discussion.---}Going beyond our scheme, it would be interesting to develop cooling protocols that explicitly satisfy detailed balance and to test whether accurate Gibbs states can be stabilized in strongly interacting fermionic systems~\cite{Chen2025, Marti2025}. It would also be interesting to explore whether quantum Mpemba-type effects~\cite{Moroder2024, Westhoff2025, Summer2025} can accelerate state preparation.

For quasi-free Lindbladian dynamics, rapid mixing and the scaling of the Lindbladian gap are known explicitly~\cite{Zhan2025}. Extending this understanding to generic strongly interacting fermionic systems requires clarifying how the Lindbladian gap and mixing time scale with system size. Matrix product state methods may help address this question in one dimension, but higher-dimensional dynamics quickly become classically intractable. In this sense, dissipative state preparation can be viewed not only as a target application, but also as a benchmark setting in which early fault-tolerant quantum devices may provide genuine advantages by directly simulating system bath dynamics beyond classical reach. 

\textit{Acknowledgements.---}We thank Mattia Moroder, Christian Kokail, Shayan Majidy, Yujie Liu, Daniel K. Mark, Tal Schwartzman, Reuben R. W. Wang, Jong Yeon Lee, Bryan K. Clark, Yizhi You and Hossein Sadeghpour for fruitful discussions. We would like to acknowledge funding from the NSF through the CUA PFC (PHY-2317134), and the DOE through the QUACQ center. H.S. acknowledges support from the NSF through a grant for ITAMP at Harvard University. L.C. acknowledges the Wellcome Leap Foundation and Harvard University for funding.

\twocolumngrid
\bibliographystyle{apsrev4-1}
\bibliography{lit}

\section*{End Matter}

\textit{Rotating wave approximation.---} In our protocol, the system and ancilla are coupled with $g (\hat{A}\otimes \hat{\sigma}^{+}+\mathrm{h.c.})$. Assume the system Hamiltonian has the following spectrum decomposition $\hat{H}_s=\sum_{k}\ket{k}\bra{k}$; then we can rewrite any traceless coupling operator as $\hat{A}=\sum_{\omega>0}\left[\hat{A}(\omega)+A^{\dagger}(\omega)\right]$ , where $\hat{A}(\omega)=\sum_{k'-k=\omega}\ket{k}\bra{k}\hat{A}\ket{k'}\bra{k'}$ is the lower energy operator. In the interaction picture, we rewrite the interaction Hamiltonian as

\eqs{
\hat{H}_{I}=& g \underset{\omega>0}{\sum}(\hat{A}(\omega)\otimes \hat{\sigma}^{+}e^{-i\Delta t}+\hat{A}^{\dagger}(\omega)\otimes \hat{\sigma}^{+}e^{i(\omega+\omega_A)t}\\ 
& +\hat{A}(\omega)\otimes \hat{\sigma}^{-}e^{-i(\omega+\omega_A)t}+\hat{A}^{\dagger}\otimes \hat{\sigma}^{-}e^{i\Delta t}).
}

In the rotating wave approximation, the second and third terms will be averaged to zero due to the fast oscillation of $\omega+\omega_A$. This will enforce energy conservation during the energy transfer process. When the interaction $g$ is weak, after the interaction time $T$, the final state is
\eqs{
\ket{\psi(T)}=&\ket{k'}\otimes\ket{0}-igT\\ & \underset{\omega=k'-k}{\sum}\langle k|\hat{A}|k'\rangle \mathrm{sinc}(\Delta T/2)e^{-i\omega T/2}\ket{k}\otimes \ket{1}\\
&+O(g^2).
}
Therefore, the probability of cooling is
\eqs{
P(k'\rightarrow k)=g^2 T^2|\langle k|\hat{A}|k'\rangle|^2\mathrm{sinc}^2(\Delta T/2)+O(g^4).
}
We see that the operator $\hat{A}$ determines the cooling channel between $\ket{k'}$ and $\ket{k}$. When $|\langle k|\hat{A}|k'\rangle|^2\neq 0$, the resonance gap $\Delta=k'-k-\omega_A$ determines the relative cooling efficiency.

\begin{figure}[t!]
\includegraphics[width=\columnwidth]{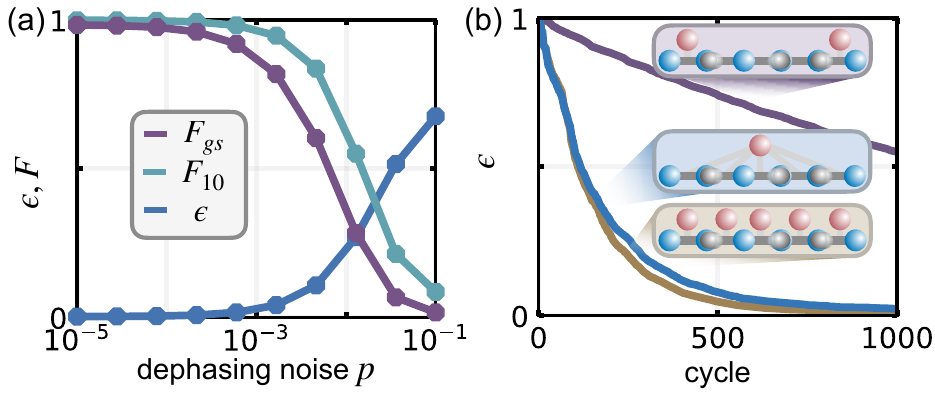}
\caption{\textbf{Noise robustness and different architectures.}
(a) Robustness against global dephasing in the 1D $t$-$t'$-$V$ model. Shown are the normalized energy $\epsilon$ and the fidelities with respect to the ground state and to the low energy subspace spanned by the ten lowest eigenstates, evaluated after 1000 cooling cycles. Parameters are identical to Fig.~\ref{fig:fig1} with $V/t = 1$. (b) Comparison of different system-ancilla architectures for a 1D chain with $V/t=4$, $L=6$, and $N_p = 3$: single global ancilla, one ancilla per bond, and boundary-only cooling.}
\label{fig:fig5}
\end{figure}

\textit{Noise robustness---}Noise resilience is a key feature of quantum algorithms for near-term devices~\cite{majidy2024building}. In the 1D $t$-$t'$-$V$ model, we apply a simple phenomenological noise model through a single site averaged global dephasing channel after tracing out the ancilla in each cooling cycle,
\begin{equation}
    \hat{\rho} \rightarrow (1-p)\hat{\rho}+\frac{p}{L}\sum_j (1-2\hat{n}_j) \hat{\rho} (1-2\hat{n}_j).
\end{equation}
Figure~\ref{fig:fig5}~(a) shows the normalized energy $\epsilon$, the ground state fidelity $F_{\text{gs}}$ as well as summed weights over the first 10 eigenstates ($F_{10}$), evaluated after $1000$ cooling cycles.

\begin{figure*}[htbp]
\includegraphics[width=\textwidth]{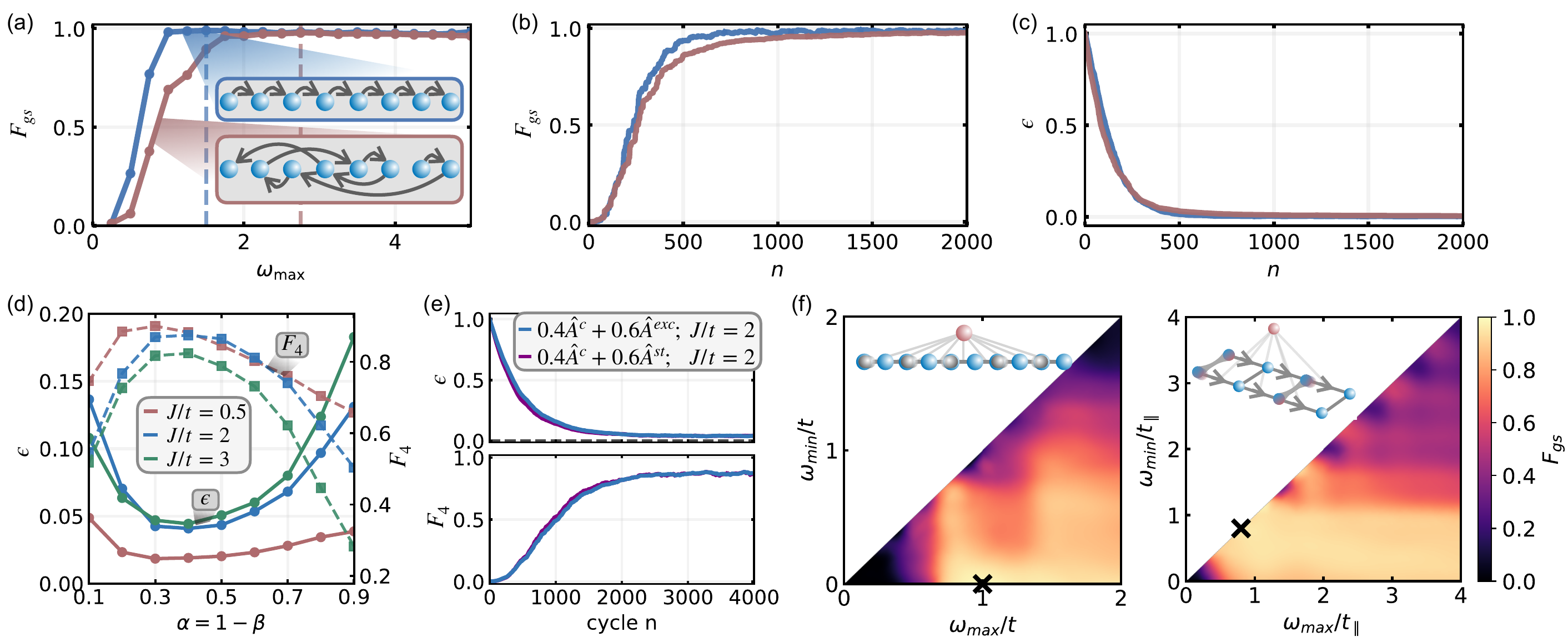}
\caption{\textbf{Additional numerical data.} (a)-(c) Comparison between directed NN and random hopping sequence in $\hat{A}$ [see inset of panel (a)] for the $t$-$t'$-$V$ model for $L=8$, $t'/t =0.5$ and $V/t = 1$. Panel (a) shows the dependence on the frequency window; the optimal choices are indicated by dashed lines. The corresponding (optimal) ground-state fidelities and normalized energies are shown in panels (b) and (c). (d) $t$-$J$ model: normalized energy density $\epsilon$ and fidelity within the four lowest energy states $F_4$ after 4000 cooling cycles for varying $J/t$ and cooling channel weights $\alpha, \beta$ (while keeping $\alpha + \beta = 1$). A broad minimum appears around $\alpha = 0.4, \beta = 0.6$. (e) Comparison between exchange spin cooling $\hat{A}^{\text{exc}}$ and singlet-triplet channel $\hat{A}^{\text{st}}$. Both channels cool comparatively, signaling that the exact choice of local cooling channel is not of fundamental importance. (f) Ground state fidelity as a function of $\omega_{\text{min}}$ and $\omega_{\text{max}}$ in the 1D $t$-$t'$-$V$ spinless fermion and mixD $t_{\parallel}$-$J_{\parallel}$-$J_{\perp}$ ladder. Black crosses show the pair $\omega_{\text{min}},\omega_{\text{max}}$ that maximizes the fidelity. Parameters are chosen as in the main text, with $V/t = 1$ in the spinless chain and $J_{\perp}/t_{\parallel}=2, J_{\parallel}/t_{\parallel} = 1$ in the mixD ladder.}
\label{fig:fig6}
\end{figure*}

Notably, for dephasing probabilities $p\lesssim 10^{-3}$ the protocol remains stable, still reaching a steady state that has close to unit fidelity. As $p$ increases further, cooling efficiency gradually degrades and a crossover occurs once the dephasing rate becomes comparable to the coherent coupling scale. The observed stability over several orders of magnitude in $p$ demonstrates that the protocol tolerates moderate dephasing and does not rely on fine tuned coherence conditions, suggesting long stabilization of interacting fermionic low-energy states of interest~\cite{Mi2024}. 

\textit{Architecture dependence.---}In Fig.~\ref{fig:fig5}~(b) we compare various system-ancilla architectures of our protocol for the 1D $t$-$V$ model with $V/t = 4$. We consider (i) a single ancilla coupled globally to all bonds, (ii) one ancilla per bond (see also Fig.~\ref{fig:fig0}), and (iii) boundary-only coupling. The first two protocols yield nearly identical cooling performance, indicating that global connectivity rather than the overall number of ancillas controls efficiency. In contrast, boundary cooling performs significantly worse, where access to bulk excitations in the interacting regime is limited.

\textit{Comparison of cooling operators.---}We compare two cooling operators for the $t$-$t'$-$V$ chain. In case (i), $\hat{A}$ is defined as a sum of directed nearest-neighbor hopping operators (as in the main text), while in case (ii) it follows a random path $(i_1 \rightarrow i_2), (i_2 \rightarrow i_3), \dots$, see Fig.~\ref{fig:fig6}~(a)-(c). We find that both choices yield comparable performance. For the random construction, however, larger values of $\omega_{\text{max}}$ are required to reach fidelities close to unity. 

This comparison further indicates that the precise microscopic structure of the coupling operator $\hat{A}$ is not crucial. As long as the operator provides sufficient overlap with downhill transitions in energy and avoids trapping the dynamics in symmetry-protected dark subspaces, the system can be cooled into a high-fidelity low-energy state.

In the main text, we have shown cooling dynamics of the $t$-$J$ model for a single ratio of spin and charge coupling channels in the unitary evolution. In particular, the coupling operator $\hat A$ was chosen as a linear combination of directed charge hopping and spin-exchange processes on nearest-neighbor and NNN bonds,
\begin{equation}
\hat A = \sum_{\langle ij\rangle, \bbrakket{ij}}\big[\alpha\sum_{\sigma}
\hat{\tilde{c}}_{i\sigma}^{\dagger}\hat{\tilde{c}}_{j\sigma}+\beta\,\hat S_i^{+}\hat S_j^{-}\big].
\label{eq:A_full}
\end{equation}

Fig.~\ref{fig:fig6}~(d) shows how a sweep over varying weights of $\alpha$ and $\beta$ (while fixing the overall strength through the constraint $\alpha+\beta=1$) leads to a broad minimum of the energy (maximum of the fidelity respectively) for values around $\alpha = 0.4$, $\beta=0.6$. This underlines that both spin- and charge channels are essential in cooling the $t$-$J$ chain, though their exact decomposition only has a minor effect on the cooling efficiency. 

We further compare results when using spin-exchange and singlet-triplet cooling in Fig.~\ref{fig:fig6}~(e), showing almost identical performance. This further suggests that the microscopic choice of the local spin cooling operators do not have a large effect on the outcome of the algorithm.  

\textit{Dependence on ancilla frequency windows.---}Lastly, we analyze the performance of the cooling algorithm as a function of $\omega_{\text{min}}, \omega_{\text{max}}$ in the 1D $t$-$t'$-$V$ model as well as the mixD $t_{\parallel}$-$J_{\parallel}$-$J_{\perp}$ ladder, Fig.~\ref{fig:fig6}~(f). We find that in the former, optimal cooling (black cross) is found when sampling $\omega$ uniformly from a range $[0,t]$, cf. Fig.~\ref{fig:fig1} in the main text. For the mixD ladder, indeed optimal cooling is found for a fixed $\omega/t_{\parallel} = 0.8$, though the whole region below $\omega_\text{min}/t_{\parallel} \lesssim 1$ yields comparable fidelities $\approx 95 \%$. 

Finally we note that, though optimal frequency windows likely depend on systems size, we expect generic early-time cooling behavior for fixed frequency windows in the thermodynamic limit, as discussed around Fig.~\ref{fig:fig1}~(d).

\end{document}